\documentclass[11pt]{article}
\usepackage{moriond,epsfig}

\def\Journal#1#2#3#4{{#1} {\bf #2}, #3 (#4)}

\def\PLB{{\em Phys. Lett.}  B}
\def\PRL{\em Phys. Rev. Lett.}
\def\PRD{{\em Phys. Rev.} D}

\def\be{\begin{equation}}
\def\ee{\end{equation}}
\def\bea{\begin{eqnarray}}
\def\eea{\end{eqnarray}}

\newcommand{\dzero}     {D\O}

\newcommand{\met}       {\mbox{$\not\!\!E_T$}}

\begin{document}
\vspace*{4cm}
\title{MESUREMENT OF TOP QUARK PROPERTIES AT THE TEVATRON}

\author{ Jessica LEV\^EQUE\\ for the CDF and \dzero ~collaborations }

\address{University of Arizona, Tucson AZ, USA 85721 }

\maketitle
\abstracts{
We highlight the most recent top quark properties measurements performed 
at the Tevatron collider by the CDF and \dzero ~ experiments. 
The data samples used for the analyses discussed correspond to an integrated luminosity 
varying from 360 pb$^{-1}$ to 760 pb$^{-1}$.    
}

\section{Introduction}
\label{intro}

\hspace{.55cm} Since its discovery at Fermilab in 1995~\cite{discover}, 
the top quark has been intensively studied but most of its properties are still poorly understood 
due to its small production cross section and the resulting statistically limited data samples. The constraints 
currently set by precision electroweak measurements leave plenty of room for exotic behaviors, and
direct measurements of top quark properties are necessary  to explore physics beyond the Standard Model. 
This paper summarizes the most recent measurements of fundamental top quark properties 
performed by the CDF and \dzero\ collaborations at the Tevatron. It covers the first direct measurement 
of the top quark lifetime and electric charge, the search for exotic production modes through 
heavy resonance decays, the search for a 4$^{th}$ generation of quarks, and the measurement 
of the $W$ boson helicity. 

At the Tevatron, the Standard Model (SM) predicts that the top quark is produced predominantly in pairs
 with a cross section of about 7 pb and to decay almost 100\% of the time into 
a $W$ boson and a $b$ quark \cite{theory}. To achieve precision measurements of top quark properties, a large and pure 
sample of top quarks is required. Therefore, only $t\bar{t}$ events where at least one of the $W$ bosons 
decays leptonically to an electron or a muon are considered. The lepton+jets and dileptons events, containing 
at least one isolated  high $P_T$ lepton, two high $P_T$ $b$ jets and a large missing transverse energy ($\met$) due to
 the escaping neutrinos in the final state represent 30\% of top quark events and 
have a relatively low background contamination. To increase the purity of the data sample, most of the analyses 
require at least one jet to be identified with a secondary vertex found inside the jet,
 which is a characteristic of the decay of long-lived particles such as $B$ or $D$ mesons.
 
\section{Top quark lifetime}

\begin{figure}
\vspace{-.5cm}
\centering
\begin{tabular}{cc}
\psfig{figure=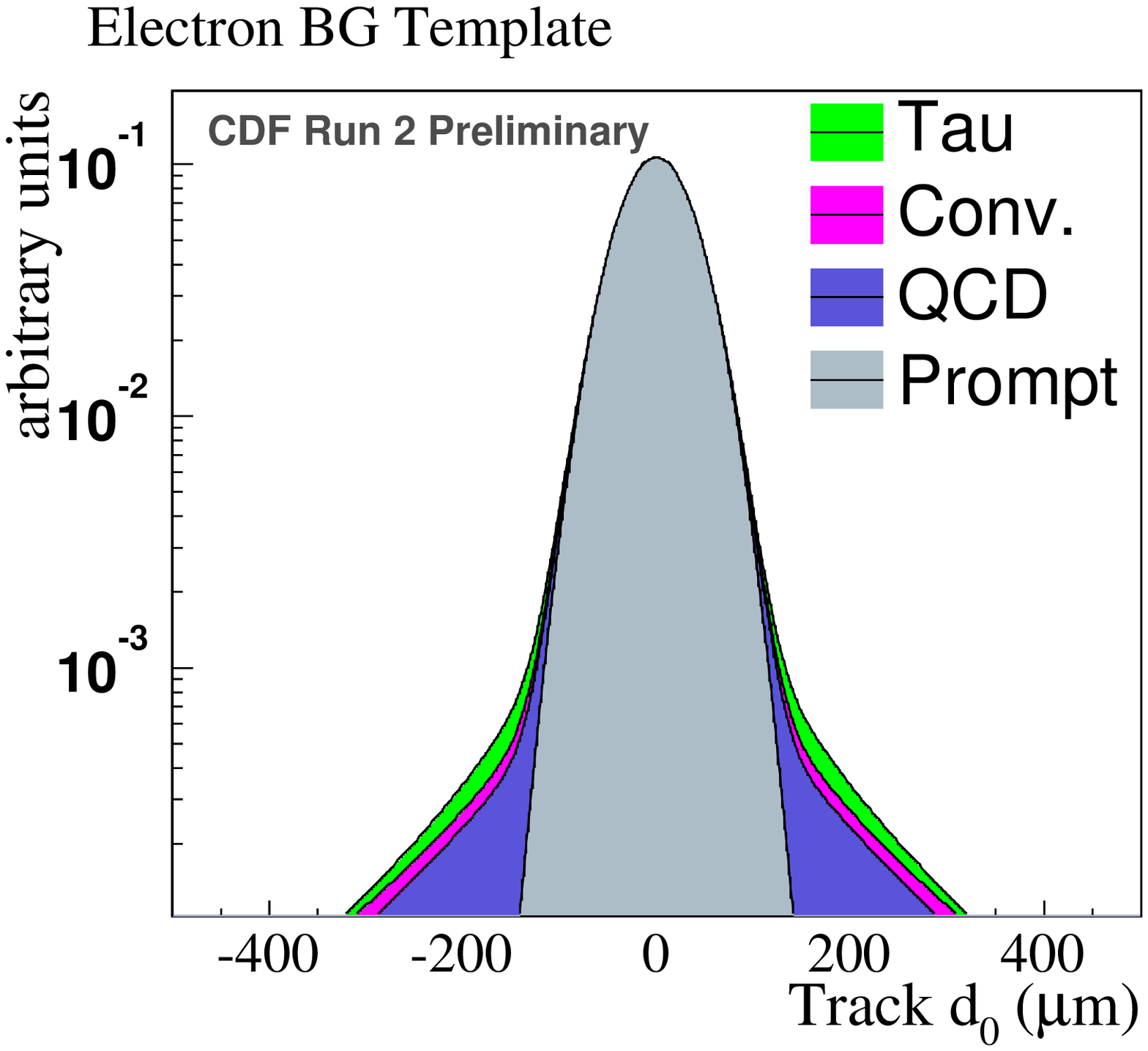,height=2.in} &
\psfig{figure=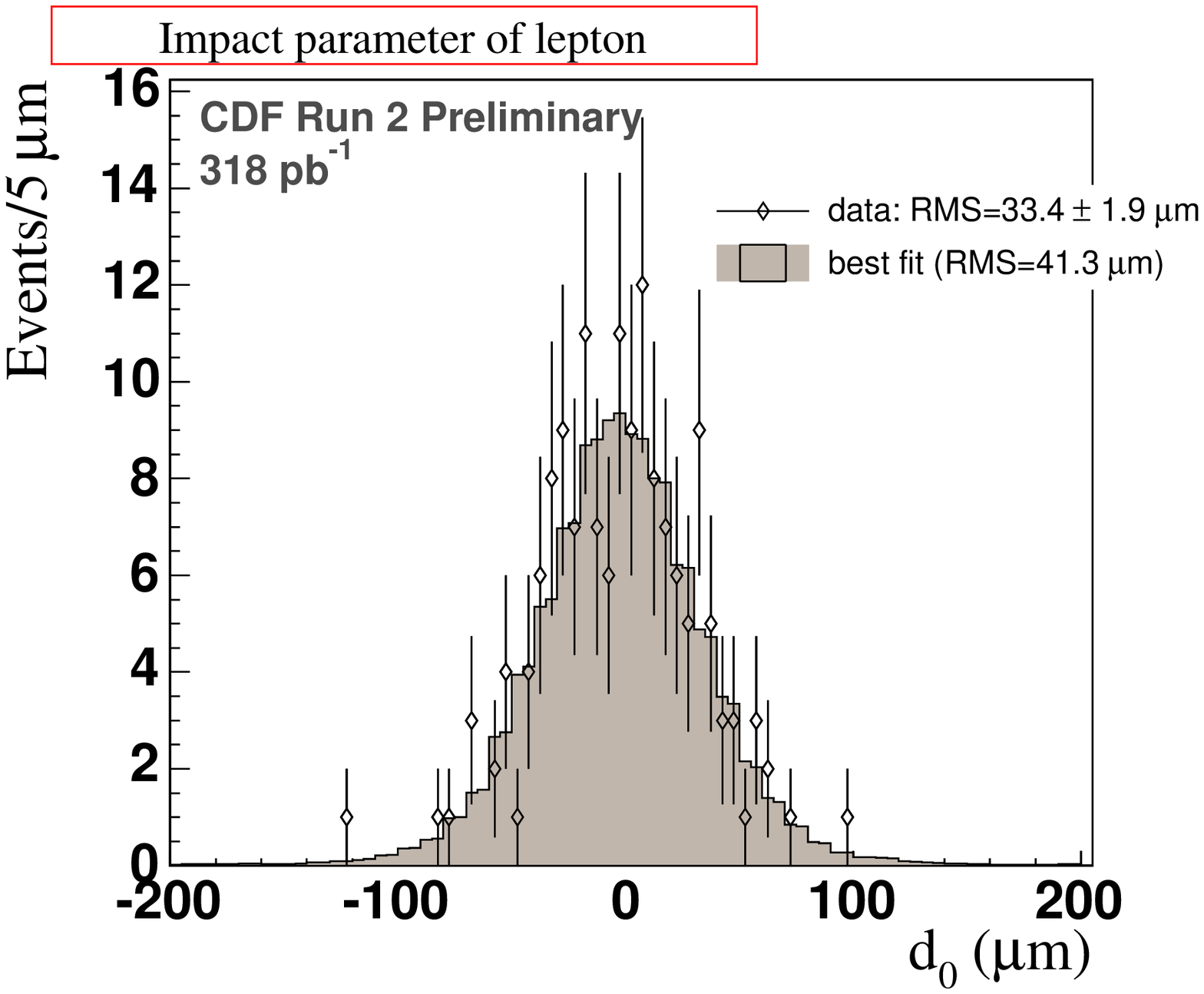,height=2.in}
\end{tabular}
\vspace{-.5cm}
\caption{Left : Lepton impact parameter ($d_{0}$) distributions of the background in the $e$+jets channel. 
Right: Comparison of the lepton $d_{0}$ distribution in the data (points) with the best 
fitted Monte-Carlo template (histogram). The histogram includes the signal $c\tau = 0~\mu m$ and 
the background contributions and is smeared to account for the detector resolution.
\label{fig1}}
\vspace{-.3cm}
\end{figure}

\hspace{.55cm} Under the SM assumptions and in the abscence of extra lepton generations~\cite{lifetime} 
the top quark lifetime is constrained to be less that 10$^{-24}s$, translating into
 a time-of-flight $c\tau < 3 \times 10^{-10}~\mu m$. The first direct measurement of the top quark lifetime 
was performed by the CDF experiment with 318 pb$^{-1}$ of data. This measurement is designed to confirm 
the identity of the top quark candidates and to detect unexpected production modes through new long-lived 
exotic particles.
The lifetime information is extracted by measuring the distance between the initial 
collision point and the W boson decay vertex, given by the lepton impact parameter $d_0$.
The lepton impact parameter distributions are built by combining the simulated top quark signal 
and backgrounds normalized to the expected sample composition.
The background consists of prompt leptons coming from $W(\to e, \mu)$+jets events, leptonic 
decays of heavy flavor jets from QCD multi-jets events, photon conversions and $W(\to \tau \to e,\mu)$+jets. 
The lepton impact parameter distributions are shown in Figure \ref{fig1} for each background component. 
The impact parameter templates are generated for a top quark lifetime varying between 
0 $\mu m < c\tau < 500~\mu m$, and smeared to take into account the detector resolution. 
The lepton $d_{0}$ distribution in data is then fitted using a maximum likelihood fit method
and is consistent with the SM prediction $c\tau = 0~\mu m$. Figure \ref{fig1} shows 
the data superimposed to the best fitted template. A limit on the top quark lifetime $ c\tau < 50 ~ \mu m$ can be derived 
  at  95\%  confidence level. 
   
\section{Top quark charge}

\begin{figure}[h]
\vspace{-.2cm}
\centering
\begin{tabular}{cc}
\epsfig{figure=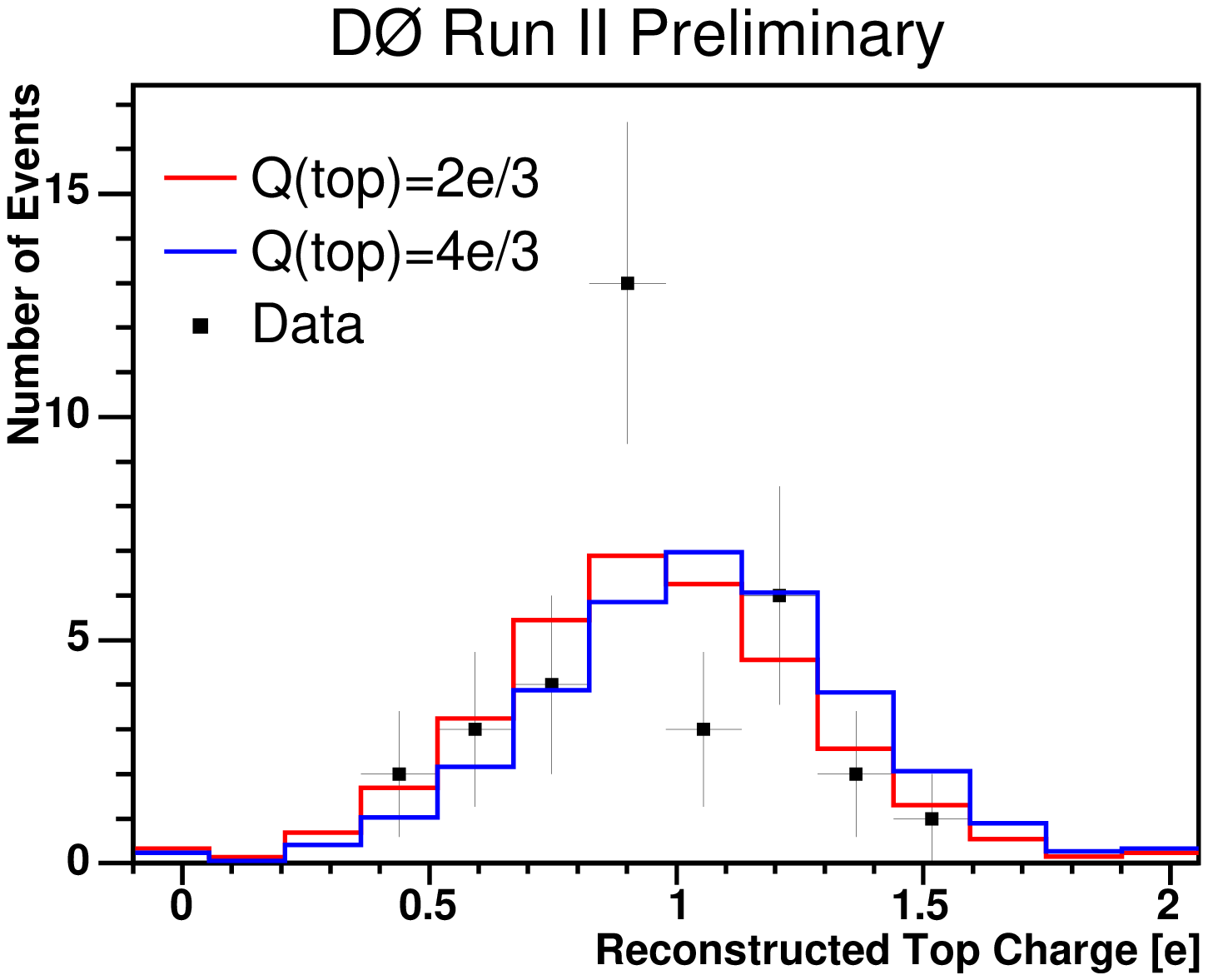,height=2.2in} &
\epsfig{figure=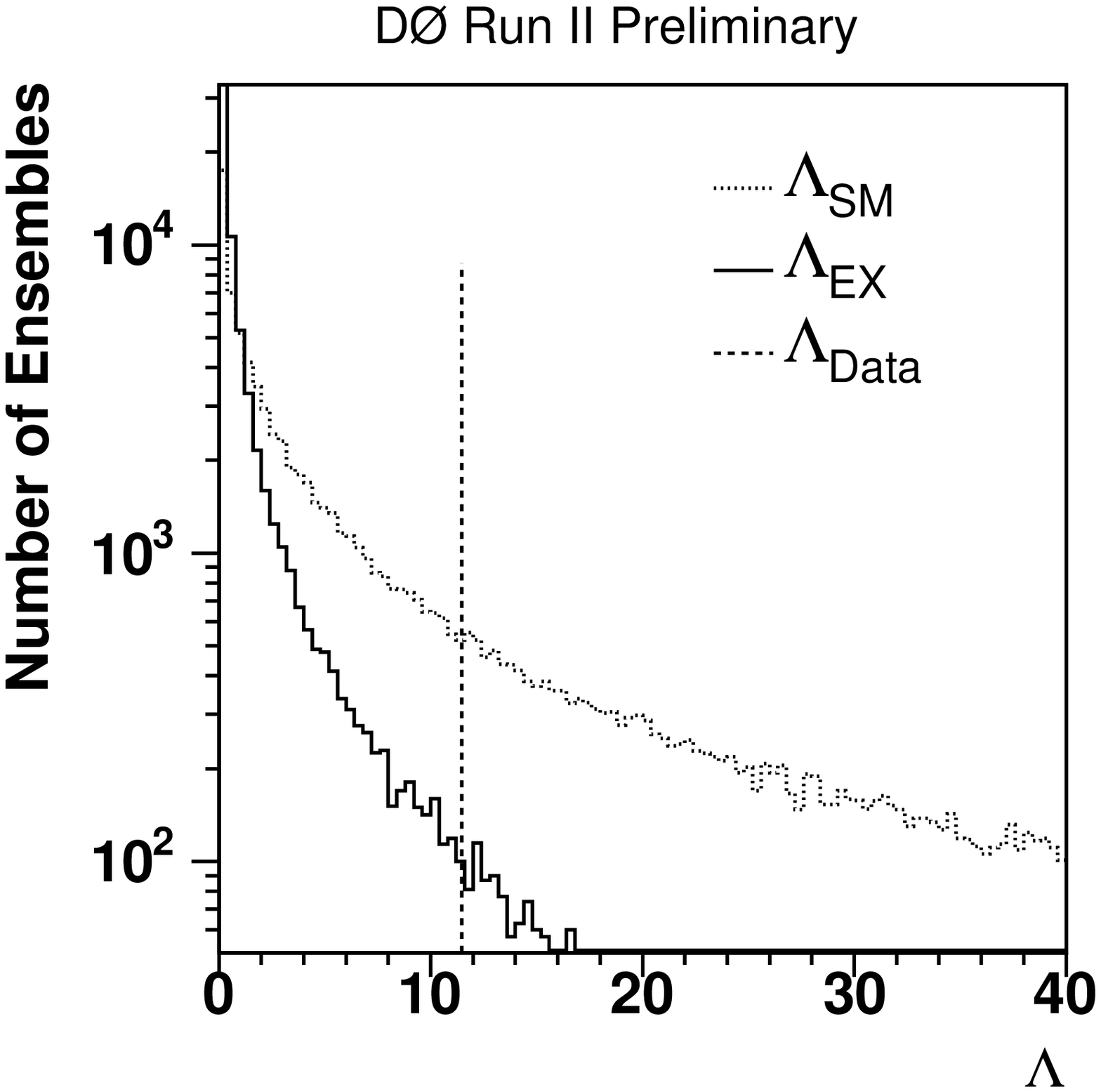,height=2.2in}
\end{tabular}
\vspace{-.5cm}
\caption{Left : Comparison of the charge distribution of the top quark 
candidates reconstructed in data (points) with the $+\frac{2}{3}e$ (red line) and $-\frac{4}{3}e$ (blue line) models.
Right : Likelihood ratio distributions $\Lambda^{SM}$ for the $+\frac{2}{3}e$ top quark charge  hypothesis (dotted line) 
and  $\Lambda^{ex}$  for the $-\frac{4}{3}e$ exotic quark charge hypothesis (solid line). 
\vspace{-.5cm}
\label{fig2}}
\end{figure}

\hspace{.55cm} In the SM, the top quark has an electric charge of $+\frac{2}{3}e$.
 However, exotic models~\cite{charge} involving a $4^{th}$ generation of heavy quarks suggest that the top quark 
candidates found at the Tevatron may be mistaken with an exotic heavy quark, very similar to the top quark 
but with an electric charge of $-\frac{4}{3}e$.  
The first direct measurement of the top quark charge has been performed by the \dzero\ experiment 
with 370 pb$^{-1}$ of data. The top (anti-top) quark charge is defined as the charge of the lepton plus (minus) 
the charge of the associated $b$ jet. The first step for this measurement requires the reconstruction of top quark
pair candidates with a kinematic fit in order to find the correct lepton/$b$ jet pairing. 
The lepton charge is determined using the curvature of the track. The $b$ jet charge 
is determined with a {\it jet charge algorithm} using all tracks with $P_{T} > $ 0.5 GeV/$c$ found in the jet to calculate
a $P_T$ weighted mean charge. The {\it jet charge} distribution 
is calibrated using an independant data sample enriched with semi-leptonic $b$ decays.
In the calibration sample, the charge of the $b$ jets is derived from the lepton charge, accounting for $c$ jet 
contamination and $B$ oscillation corrections. Figure \ref{fig2} compares the charge distribution of the top quark 
candidates reconstructed in data with the $+\frac{2}{3}e$ and $-\frac{4}{3}e$ charge hypotheses. 
A likelihood ratio defined as $$\Lambda = \frac{\Pi_{i}p^{sm}(q_{i})}{\Pi_{i}p^{ex}(q_{i})}$$ is then computed to determine 
the most probable model. The numerator measures the probability $p^{sm}$ for a charge distribution to agree with the 
SM hypothesis, while the denominator measures the probability $p^{ex}$ for the charge distribution 
to agree with the exotic quark model. The value of the likelihood ratio measured in data is compared 
to the $\Lambda^{SM}$ and $\Lambda^{ex}$ distributions, derived by performing ensemble tests
using the SM or the exotic quark charge hypothesis for the signal respectively. For the exotic heavy quark hypothesis, 
only 6\% of the pseudo-experiments gives a higher $\Lambda$ ratio than the one measured in data (Figure \ref{fig2}). 
Therefore, the existence of heavy quarks with a charge $-\frac{4}{3}e$ can be excluded at 94\% confidence level.

\section{Search for $t\bar{t}$ resonant production}

\hspace{.55cm} This search was performed by both CDF and \dzero\ collaborations and is dedicated to testing 
the hypothesis of  top quark pair-production through the decay of a new heavy gauge boson. In such a case,
the measured production cross section would be higher than predicted by the SM and a 
resonance would appear in the $t\bar{t}$ invariant mass distribution. The search is model-dependant \cite{resonance},  
and restricts the resonance mass $M_{X}$ to the range [350 - 1000] GeV/$c^{2}$ under the assumption of
a width $\Gamma_{X} = 0.012 M_{X}$. The jet/lepton assignement the most consistent with the top decay hypothesis is 
chosen and the invariant mass of the reconstructed top quark pair candidates found in data is compared 
to the background and top quark distributions, normalized to the SM expectation.
 Since no excess is observed in data, both experiments can exclude a leptophobic $Z'$ boson with 
a mass $M_{x} < 680$ GeV/$c^{2}$ at 95\% confidence level for \dzero\ with 370 pb$^{-1}$ of data 
and  $M_{x} < 725$ GeV/$c^{2}$ at 95\% confidence level for CDF with 682 pb$^{-1}$ of data.

\section{Search for new heavy quarks $t'$ }

\hspace{.55cm} This search was performed by the CDF collaboration with 760 pb$^{-1}$ of data under the assumption 
that the exotic quark is produced in pairs, decays promptly into a $W$ boson and a $b$ jet, and has a higher mass 
than the SM top quark \cite{tprime}. In order to distinguish the $t'$ signal from the top quark, 
a 2-dimensional function using the reconstructed mass of the heavy quark candidates and 
the total transverse energy in the event is built. The 2-dimensional distribution is then fitted with a binned likelihood 
method using a combination of templates for the $t'$, $t\bar{t}$, $W$+jets and other SM backgrounds contributions.
The $W$+jets production cross-section is a free parameter of the fit, while the $t\bar{t}$ and other backgrounds 
contributions are normalized to the SM expectation. Signal templates with $M_{t'}$ varying in the 
[175 - 400] GeV/$c^{2}$ range were considered, but no evidence of a $t'$ signal was found. 
Therefore heavy quarks with a mass 
$M_{t'} < $ 258 GeV/$c^{2}$ can be excluded at 95\% confidence level. 

\section{W boson helicity}
    
\begin{figure}[h]
\centering
\begin{tabular}{cc}
\epsfig{figure=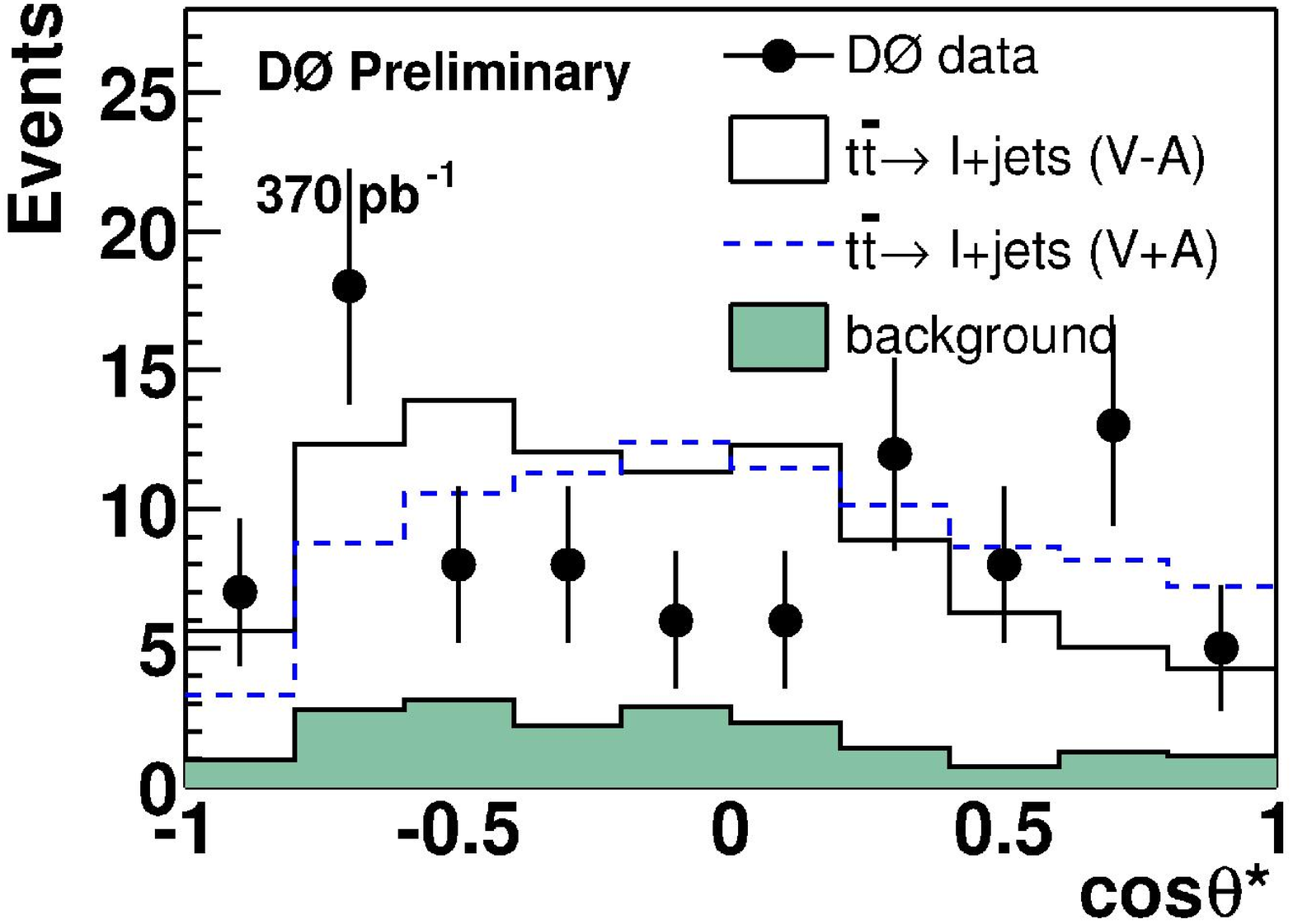,height=1.8in} &
\epsfig{figure=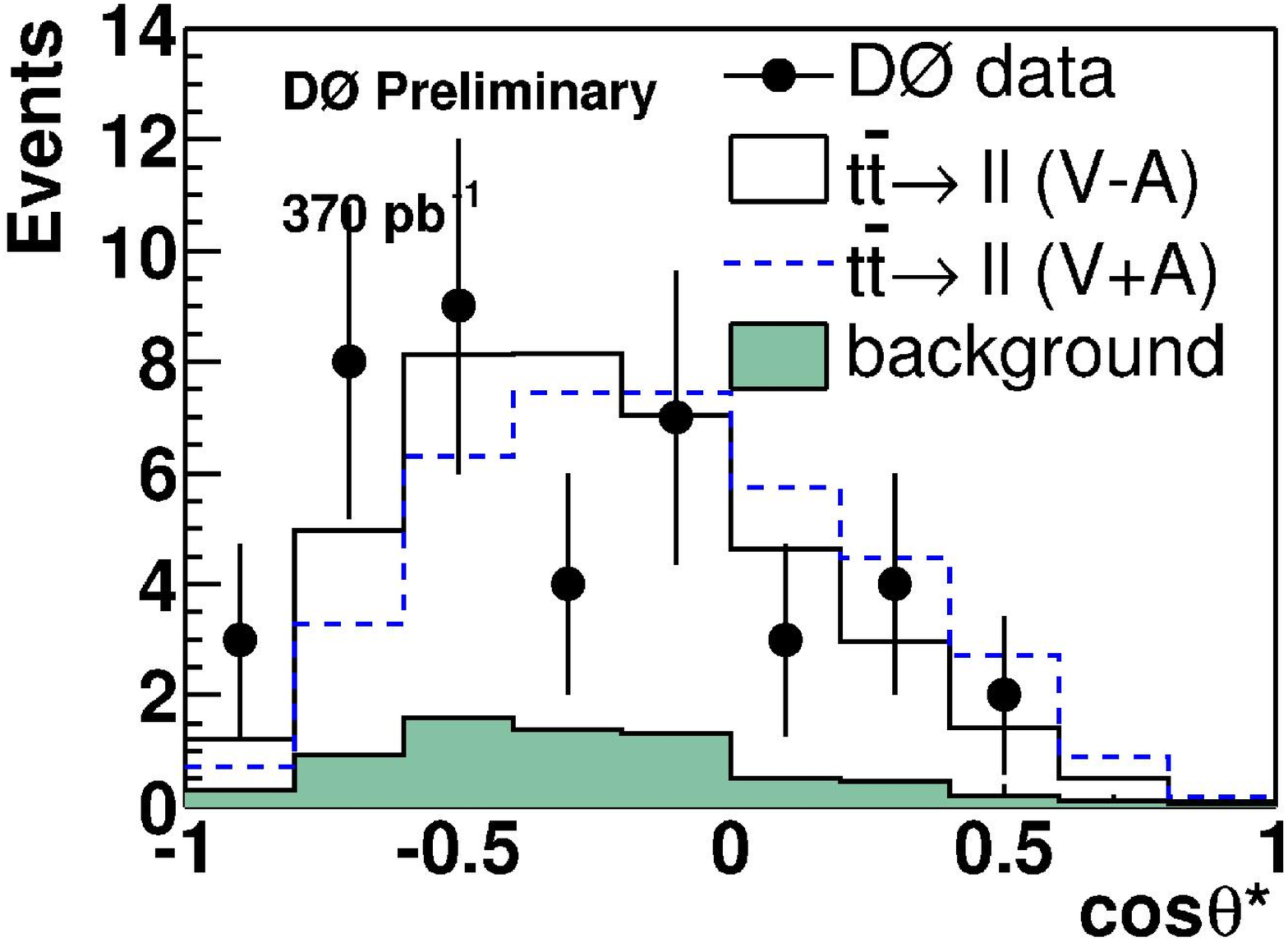,height=1.8in}
\end{tabular}
\vspace{-.5cm}
\caption{Comparison of the $cos~\theta^{*}$ distribution 
in data (point) with the Monte-Carlo distributions expected for the pure V-A and pure V+A couplings hypotheses (histograms)
in lepton+jets (left) and dilepton (right) channels.
\label{fig4}}
\vspace{-.5cm}
\end{figure}

\hspace{.55cm}  The analysis presented here was done by the \dzero\ collaboration and used 370 pb$^{-1}$ of data. 
Assuming a massless bottom quark, the V-A structure of the SM weak interaction constrains 
the fraction of $W$ bosons with right helicity ($f_{+}$) produced in $t\to Wb$ decay to be less than $3.6 \times 10^{-4}$.
The fractions of $W$ bosons with longitudinal ($f_{0}$) and left ($f_{-}$) helicities  are 
expected to be 70\% and 30\% respectively.  Experimentally, the $W$ boson helicity is measured from 
the angle $cos~\theta^{*}$ between the charged lepton and the top quark boost directions, 
measured in the $W$ boson rest frame. Leptons coming from right-handed $W$ boson are 
preferentially emitted along the top quark direction, leading to smaller  $cos~\theta^{*}$ values 
than for left-handed $W$ bosons.  A kinematic fit is used to reconstruct the top quark pairs and the $W$ boson rest-frame. 
In dilepton events, the presence of two neutrinos leads to an underconstrained system and gives 
two possible angle measurements. Both combinations show discrimination between the V-A and V+A models and 
are used in the fit. Figure \ref{fig4} compares the $cos~\theta^{*}$ distribution 
in data with the simulated distributions for the pure V-A and pure V+A couplings hypothesis in
lepton+jets and dilepton channels. The measured value of $cos~\theta^{*}$ is extracted using a binned maximum likelihood fit 
for each channel. The combined result gives $f_{+} = 0.06 \pm 0.08 (stat) \pm 0.06 (syst)$ and allows for the setting of
an upper limit of $f_{+} < 0.23$ at 95\% confidence level.
 
\section{Conclusion}

\hspace{.55cm} The top properties measured so far are consistent within errors with SM predictions.
The analyses are still statistically limited, but were intensively optimized and have already reached interesting precisions. 
The CDF and \dzero\ collaborations have now more than 1 fb$^{-1}$ of data recorded. These data are currently being analyzed.  
and more precise measurements are expected for the summer of 2006.

\end{document}